# Nonlinear Frequency Response of the Multi-resonant Ring Cavities

Andrey A. Nikitin[a,*], Vitalii V. Vitko[a], Mikhail A. Cherkasskii[a,b], Alexey B. Ustinov[a], and Boris A. Kalinikos[a]

[a]St. Petersburg Electrotechnical University 'LETI', Department of Physical Electronics & Technology, St. Petersburg, 197376, Russia
[b]St. Petersburg State University, Department of General Physics 1, St. Petersburg, 199034, Russia
*and.a.nikitin@gmail.com

**Abstract**

A convenient for practical use new theoretical approach describing a nonlinear frequency response of the multi-resonant nonlinear ring cavities (RC) to an intense monochromatic wave action is developed. The approach closely relates the many-valuednesses of the RC frequency response and the dispersion relation of a waveguide, from which the cavity is produced. Arising of the multistability regime in the nonlinear RC is treated. The threshold and the dynamic range of the bistability and tristability regimes for an optical ring cavity with the Kerr nonlinearity are derived and discussed.

1. **Introduction**

The resonant ring cavities (RC) belong to the core constructions of modern technology. They are utilized to produce various passive and active devices of optics, microwave electronics, spin-wave electronics, plasmas, and other (see, e.g. [1-4]). As is well known, two conditions are necessary to observe a resonant behavior. The first one is a phase condition. It consists in an in-phase addition of all the waves circulating in the ring. The second one is a dissipative condition. It demands a small attenuation, which is necessary for a multiple addition of the circulating waves. It is the multiple in-phase addition that leads to the resonant enhancement of the signal at a certain frequency, in case the both conditions are satisfied. The resonant enhancement provides a decrease in the nonlinear processes threshold, which makes the RC exceptionally useful constructions to study a variety of the nonlinear phenomena. One of them is a dispersive bistability [5]. The dispersive bistability effect manifests itself in a single RC as a frequency shift of each resonant harmonic with increasing of the input signal intensity. Under certain intensity corresponding to the tristability threshold, the multiple tilted resonances begin overlapping and form a tristability phenomenon. In order to reduce the threshold it is necessary to bring the neighboring resonant frequencies closer to each other. For a single RC it is a contradictory requirement lying in the fact that an increase of the ring dimensions reduces its free spectral range and enhances a bistability threshold intensity [6-9]. The multi-ring resonant systems allow for resolving this demand and hence provide a substantial decrease in the tristability threshold [10-12].

Nowadays the Lugiato-Lefever equation (LLE) is widely used to describe a wide variety of nonlinear phenomena existing in the ring cavities [13-18]. This equation was derived using both the mean-field approach and infinite-dimensional map (Ikeda map) [19-21]. Recently the LLE was extended to consider the RC multiple nonlinear resonances as well as the multi-valued stationary states [22,23]. In parallel, such states were independently investigated with a general Ikeda map [24]. The investigations done beyond the conventional LLE open new possibilities to describe the multistability in the solitary and coupled micro-rings [25-27].

The complexity of the problem devoted to the analysis of nonlinear wave processes in multi-ring system rises manifold with increasing RC number and changing the geometry of the resonance system. The aim of this paper is to propose a new convenient for practical use theoretical approach, which bonds the effects of the nonlinear dispersion and frequency response of the multi-resonant nonlinear ring cavities. The approach enables one to relate the many-valuednesses of the RC frequency response and the dispersion characteristic of a waveguide the RC is made of. Being the field of the extended LLE, our theoretical approach does not challenge the existing literature, yet, at the same time, it introduces a new essential feature. This feature consists in a self-consistent unification of the multi-ring frequency response and changing in the dispersion law. It is the feature that differs our approach from the known ones.

2. **Theoretical approach to describe nonlinear multi-resonant RC**

Let us consider a multi-resonant RC of the length $l$, which includes of a unidirectional coupler placed between a ring resonator and a waveguide shown (see Fig. 1). Assume that a plane monochromatic wave with amplitude $A_{in}$ is applied to the RC input. The RC coupling with the drop-in and drop-out waveguides is described by the power coupling coefficients $\kappa_1$ and $\kappa_2$, respectively. On assumption of the lossless coupling, the amplitude of the wave applied to the ring after the drop-in coupler is $A_0^{in} = \sqrt{\kappa_1} A_{in}$. Every single round trip over the ring is described by the wave factor $T = \sqrt{1-\kappa_1}\sqrt{1-\kappa_2} \exp(i\beta l - \alpha l)$ where $\beta$ is a propagation constant and $\alpha$ is a damping decrement. Denoting the resulted amplitude of the circulating signal by $A_c$, we find it as a superposition of an infinite number of the circulating waves in the form $A_c = A_0^{in} \sum_{q=1}^{\infty} T^q$ where $q$ is a summation index representing a number of the wave circulations. As far as the damping decrement in a dissipative medium has a positive value $(\alpha > 0)$, the series converges and after summation the resulted complex amplitude becomes

$$A_c = \frac{A_0^{in} T}{1-T}. \tag{1}$$

As far as the intracavity intensity has the following form $I_c = |A_c|^2$, it reads

$$I_c = \frac{I_0 \kappa_1 (1-\kappa_1)(1-\kappa_2)}{e^{2\alpha l} - 2e^{\alpha l}\sqrt{(1-\kappa_1)(1-\kappa_2)}\cos(\beta l) + (1-\kappa_1)(1-\kappa_2)} \tag{2}$$

where $I_0 = |A_{in}|^2$ is the input intensity.

The output signal results in a superposition of circulating waves $A_{out} = A_0^{out} \sum_{q=0}^{\infty} T^q$ where $A_0^{out} = A_{in}\sqrt{\kappa_1}\sqrt{\kappa_2} \exp(i\beta l_x - \alpha l_x)$ represents the wave amplitude after propagation trough a section of the ring from the drop-in to the drop-out couplers with the length $l_x$ (see orange dots in Fig. 1). Defining the complex transmission coefficient as $H = A_{out}/A_{in}$, we obtain the frequency response of the RC as

$$H_p = |H|^2 = \frac{\kappa_1 \kappa_2 e^{2\alpha(l-l_x)}}{e^{2\alpha l} - 2e^{\alpha l}\sqrt{(1-\kappa_1)(1-\kappa_2)}\cos(\beta l) + (1-\kappa_1)(1-\kappa_2)} \tag{3}$$

The derived relation demonstrates a multi-resonant behavior because of the constructive interference of the circulating waves having the resonant wave-numbers $\beta_m = 2\pi m/l$ where $m$ is a mode number. The corresponding frequency response is found by substitution of the appropriate dispersion relation $\beta(\omega)$ and coupling coefficients $\kappa_1(\omega)$ and $\kappa_2(\omega)$ in Eq. (3).

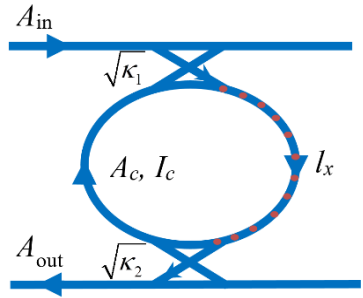

Fig. 1. Schematic of resonant ring cavity.

Note that relations akin Eq. (3) were used in a number of investigations to describe the frequency response of the optical ring resonators [28, 29], the spin-wave and multiferroic active ring resonators [2,30], the optoelectronic rings [31], and the spin-wave optoelectronic rings [32–35]. A critical feature of the frequency response Eq. (3) is an enhancement of the intracavity field intensity close to the resonant frequencies, being the effect to take into account for the multi-resonant rings. To study this effect thoroughly, one should consider the dependence of an RC performance on the wave sensitivity of the waveguide parameters of which it is made.

## 3. Application of the theoretical approach to a nonlinear optical ring

In this work, we use an optical wave intensity dependent refractive index as the nonlinear parameter of the waveguide substance. In the case of the Kerr nonlinearity, we write $n(I) = n_0 + n_2 I$ where $n_0$ is a linear refractive index, $n_2$ is a second-order refractive index, and $I$ is an optical wave intensity. For investigation of the nonlinear RC, we substitute the intracavity intensity $I_c(\omega)$ obtained from Eq. (2) into the expression for refractive index, getting $n(\omega, I_0) = n_0 + n_2 I_c(\omega, I_0)$. The latter formula shows that the refractive index is enhanced at all the resonant frequencies.

The next step in introducing our approach should be done utilizing some dispersion law for a regular waveguide fabricating the resonant RC. For demonstration of a generality and advantages of our approach, we will use the simplest approximation for the linear dispersion law

$$\omega = c\beta/n \qquad (4)$$

where $c$ is the speed of light. Note that such an approximation is valid for the frequency band, in which the dispersion impact on the wave process is relatively weak. Usually it works well for analysis of several neighboring resonant frequencies. Provided it is necessary to study a nonlinear RC in a wide frequency range, one should consider a higher-order dispersion.

To study the nonlinear effects, we substitute the intensity dependent refractive index $n(\omega, I_0)$ instead of the linear one in the linear dispersion law, given with Eq. (4):

$$\omega(\beta, I_0) = \frac{c\beta}{n_0 + \dfrac{n_2 \kappa_1 (1-\kappa_1)(1-\kappa_2) I_0}{\exp(2\alpha l) - 2\exp(\alpha l)\sqrt{(1-\kappa_1)(1-\kappa_2)} \cos(\beta l) + (1-\kappa_1)(1-\kappa_2)}}. \qquad (5)$$

Thus, we arrived at the nonlinear dispersion relation. Using this relation, it is straightforward finding the nonlinear transmission coefficient/response of the RC $H_p(\omega, I_0)$. We find it through substitution of a solution of Eq. (5) into Eq. (3). It is important to underline that the obtained nonlinear relations for $\beta(\omega, I_0)$ and $H_p(\omega, I_0)$ are the functions of two-variables. For their demonstrative description, it is instructive to introduce the dispersion surfaces. Here and after, we will name them as the nonlinear dispersion surfaces and the nonlinear response surfaces. Fig. 2.a and 2.b show the color-mapped images of the both surfaces.

In the calculations we use parameters typical for the silicon nitride RC having the radius of 300 μm. To summarize the obtained results, we constructed the figures using following notations: the dimensionless frequency $\Omega = (\omega - \omega_m)/\Delta\Omega$ where $\omega_m$ is a resonant frequency with number $m$, the free spectral range $\Delta\Omega/2\pi$, the dimensionless wave-number $\mathrm{B} = (\beta - \beta_m)/\beta_m$ where $\beta_m = 2\pi m/l$ is a resonant wave-number, the dimensionless frequency response $\mathrm{T} = H_p(\omega, I_0)/H_p(\omega_m, 0)$ where $H_p(\omega_m, 0)$ is a linear frequency response at the resonant frequency. On the base of the obtained data, it is possible to do valuable conclusions on the nonlinear resonant RC characteristics. Below we discuss some of them only.

From the figures, one can see a fascinating behavior of the nonlinear RC characteristics that manifest themselves with increasing in the input optical intensity $I_0$. The green lines calculated for the low input intensity ($I_0 = 0.1 I_{th}^{II}$) show the linear dispersion characteristic and frequency response. An increase of the input intensity provides increasing in the refractive index, which produces downshift of all the resonant frequencies. This effect is enhanced due to the constructive interference of circulating waves and an increase of the intracavity intensity in vicinity of the RC resonant frequencies. Moreover, our approach allows one to relate the nonlinear dispersion shift and the power of the signal circulating in the ring.

The blue lines in Fig. 2.a and 2.b present the characteristics calculated for the threshold intensity ($I_0 = I_{th}^{II}$). The threshold intensity is defined as maximum intensity of the input signal, for which a single solution of the two equations, Eq. (3) and Eq. (5), exists. The expression for the threshold intensity will be obtained and discussed below. Note that an increase in the input intensity $I_0$ higher than $I_{th}^{II}$ provides appearance of the region with two stable and one unstable state of the intracavity intensity, which corresponds to the *bistability phenomenon*. The magenta lines in the both figures show the dispersion characteristics and the frequency response for the intensities higher than the threshold value ($I_0 = 15 I_{th}^{II}$). Further increase in the input intensity (up to $I_0 = 150 I_{th}^{II}$) extends the frequency bandwidth of the instable behavior. Our calculations show that the new unstable states can develop progressively with increasing of the circulating power. Such a behavior corresponds to the multistability regime (see orange lines in Fig 2.a and 2.b). So, for the nonlinear frequency shift ($I_0 = 150 I_{th}^{II}$), which is more than one free spectral range ($\Delta\Omega > 1$), three stable and two unstable values of the intracavity intensity and corresponded wave-number multivaluedness appear. Furthermore, more and more unstable states may advance with increasing of the circulating power.

It is clear physically that in order to have multistability, it is necessary to observe the conditions under which the nonlinear frequency shift is more than the distance between two adjacent frequencies. In case of a single ring, this requires a sufficiently high input power, which in the real-life resonant ring cavities cannot be achieved due to a nonlinear damping. However, the multistability threshold can be significantly reduced in some cases like coupled ring systems [11]. This effect will be discussed in details below.

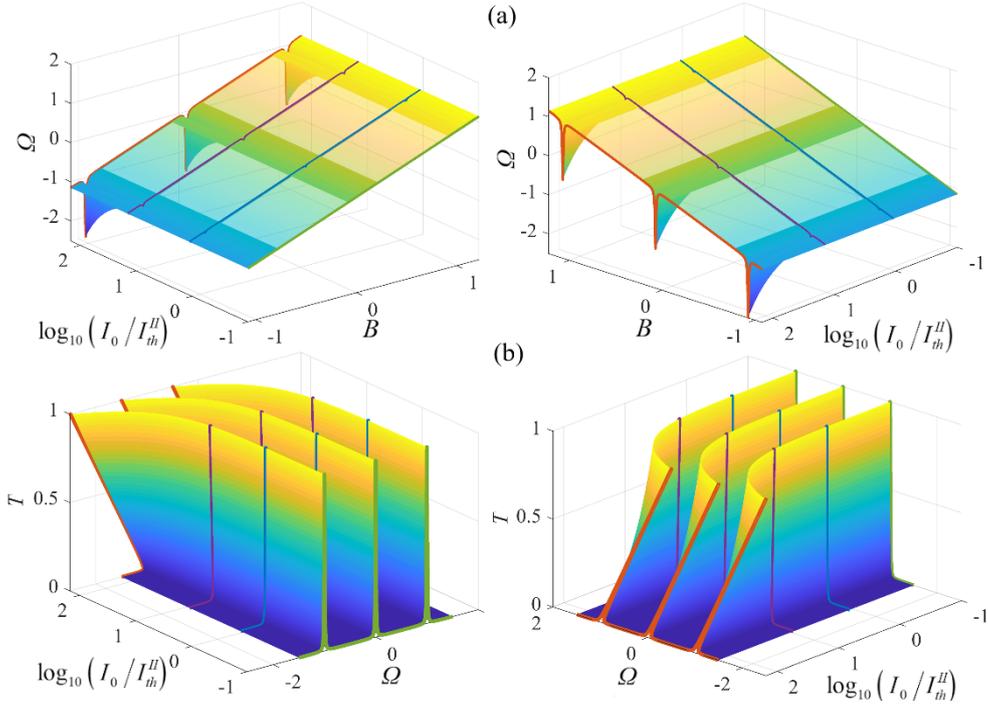

Fig. 2. Color-mapped nonlinear dispersion surfaces for a resonant RC calculated for the various intensities and shown for the different viewing angles (a), color-mapped nonlinear response surfaces for a resonant RC calculated for the various intensities and shown for the different viewing angles (b).

## 4. Optical ring bistability threshold

As an example, we consider the bistability phenomenon in an optical ring cavity. As was already mentioned, one of the distinctive features of the bistability behavior is an appearance of the two-valuedness in the relationship between the intracavity and input intensities. Fig. 3 shows the nonlinear dispersion characteristics and the frequency responses calculated with the developed approach using Eq. (5) and Eq. (3) for the various input intensities $I_0$. The green solid lines calculated for zero input intensity represent the linear dependences. An increase in the intensity up to the bistability threshold causes appearance of the kinks $\partial \omega(\beta, I_{th}^{II})/\partial \beta = 0$ and $\partial H_p(\omega, I_{th}^{II})/\partial \omega = \infty$ on the dispersion characteristic and frequency response (see the orange dotted lines in Fig. 3.a and 3.b). As a help for eyes, these points are shown on the curves with the open black squares.

Due to the nonlinearity, the dispersion relations $\omega(\beta)$ and the frequency response $H_p(\omega)$ become multi-valued slightly above the threshold value (see magenta dash-dotted lines for $I_0 = 3I_{th}^{II}$ and black dashed lines for $I_0 = 6I_{th}^{II}$ in Fig 3.a and 3.b). The bistability region corresponds to the frequency range bounded by the extremums shown in Fig. 3 by the orange and blue open circles. Note that these extremums correspond to the group velocity zeros for the wave under consideration. As is seen from Fig. 3.a, the nonlinear dispersion characteristics coincide with the linear one in the frequency range situated far enough from the given resonance.

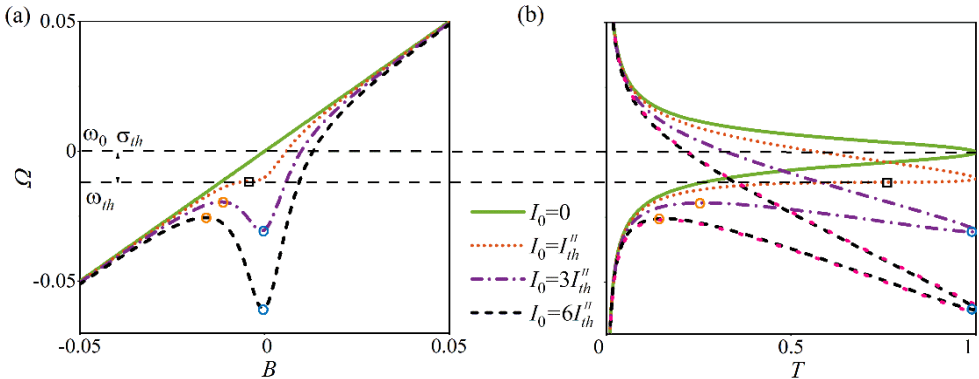

Fig. 3. Fragments of the nonlinear dispersion characteristic (a) and the frequency response (b) for the various input intensities.

For a comparison, the nonlinear frequency response for the maximum input intensity $(I_0 = 6I_{th})$ was also calculated for a critically coupled RC using classical approach for a nonlinear resonance (see red dotted line in Fig. 3.b) described by:

$$\omega = \omega_m - \mu I \pm \sqrt{\frac{F^2}{I} - \delta^2} \qquad (6)$$

where $\mu I = c\beta_m n_2 I/n_0^2$ is a nonlinear frequency shift, $F = \sqrt{\alpha c^2 I_0/(n_0^2 l)}$ is an external driving force, $\delta = 2\alpha c/n_0$ describes linear losses [36]. One can see that obtained results are in a good agreement.

Turn now to the bistability threshold. We stress that Eq. (3) and Eq. (5) describe nonlinear properties of the RCs in a wide frequency range, which includes a set of resonant modes. Both bistability and the multistability threshold for one or more resonant modes can be found numerically from these equations, while the Eq. (6) is applicable for each bistable regime individually. Let us demonstrate explicitly the simplifications that are introduced in our approach to find an analytical description of the bistability phenomenon. We rewrite dispersion relation Eq. (4) for the mode with an arbitrary number $m$ in the following form:

$$\omega = \frac{c}{n}(\Delta\beta + \beta_m) \qquad (7)$$

where $\Delta\beta = \beta - \beta_m$. For definiteness, we further consider a critically coupled RC, which under assumption of the identical coupling $\kappa_1 = \kappa_2 \equiv \kappa$ leads to $(1-\kappa)^2 = \exp(-2\alpha l)$. Following the developed approach, we substitute the intensity dependent refractive index $n(\omega, I_0)$ instead of the linear one in Eq. (7):

$$\omega(\beta, I_0) = \frac{c(\Delta\beta + \beta_m)}{n_0 + \dfrac{n_2(1-\exp(-\alpha_c l/2))I_0}{\exp(2\alpha_c l) - 2\exp(\alpha_c l)\cos(\Delta\beta l) + 1}}. \qquad (8)$$

Here $\alpha_c = 2\alpha$ is a generalized damping decrement for the critically coupled RC. Further simplifications are introduced by using the Maclaurin's expansions for $\cos(\Delta\beta l) \cong 1 - \Delta\beta^2 l^2/2$ and $\exp(\pm\alpha_c l) \cong 1 \pm \alpha_c l$. Using these expansions and neglecting the third and higher order terms, one obtains Eq. (8) as

$$\omega(\beta, I_0) = \frac{c(\Delta\beta + \beta_m)}{n_0 + \dfrac{n_2 I_0}{2\alpha_c l\left(1 + \dfrac{\Delta\beta^2}{\alpha_c^2}\right)}}. \qquad (9)$$

We underline that the made simplifications define the applicability boundaries for Eq. 9. The latter equation is valid for small detuning from the resonant frequency $(|\Delta\beta|l \ll 1)$, small losses $(\alpha_c l \ll 1)$, and weak dispersion. The last assumption is caused by using the linear dispersion law described by Eq. (7). As was already mentioned above, the bistability threshold manifests itself as the kink on the dispersion surface $\partial\omega(\beta, I_{th}^{II})/\partial\beta = 0$. This condition is satisfied when a numerator of the derivative is equal zero. Denoting the numerator by $V$, one obtains

$$V = \Delta\beta^4 + C_2\Delta\beta^2 + C_1\Delta\beta + C_0 \qquad (10)$$

where $C_2 = \dfrac{\alpha_c}{2n_0 l}(4\alpha_c n_0 l + 3I_0 n_2)$, $C_1 = \dfrac{\alpha_c}{n_0 l}\beta_m I_0 n_2$, and $C_0 = \dfrac{\alpha_c^3}{2n_0 l}(2\alpha_c n_0 l + I_0 n_2)$.

Equation (10) is equivalent to the one used in the cusp catastrophe [37]. The number of the real roots of Eq. (10) is defined by the value of the input intensity $I_0$. Eq. (10) has a single real root in the case when its discriminant vanishing $D = 0$. As in the catastrophe theory, the threshold is found from the conditions $V = 0$ and $D = 0$. These conditions give the threshold value, which reads as

$$I_{th}^{II} = \frac{8\sqrt{3}}{9}\frac{\pi n_0^2 l}{Q^2 n_2 \lambda} \qquad (11)$$

where $Q = \beta/2\alpha_c$ is a quality factor of the critically coupled RC in the linear regime, $\lambda$ is a free-space wavelength of the laser radiation.

Substitution of Eq. (11) in Eq. (10) finds the threshold value of the nonlinear wavenumber shift $\Delta\beta_{th} = -\alpha/\sqrt{3}$. Substituting of the last expression to the dispersion relation (7) provides the threshold value of the frequency shift $\sigma_{th} = -\sqrt{3}\Delta\omega/2$. Here, a notation $\sigma_{th} = \omega_{th} - \omega_m$ characterizes an offset of the resonant frequency $\omega_m$ due the input intensity $I_0 = I_{th}^{II}$ and notation $\Delta\omega = 2c\alpha_c/n$ describes a half-power bandwidth. It is clear that if both the frequency shift and input intensity exceed the threshold values $\sigma > \sigma_{th}$ and $I_0 > I_{th}^{II}$ the bistability should appear. The derived expressions for the threshold intensity and the nonlinear frequency shift coincide with the results obtained in a series of investigations devoted to the Kerr nonlinearity [6-9].

## 5. Dynamic range of the bistability phenomenon

For analysis of the bistability dynamic range, we use the diagram of the nonlinear behavior [37]. As in the catastrophe theory, such diagram is a curvilinear surface, which demonstrates all real roots of Eq. (10). The diagrams presented in Fig. 4 show nonlinear behavior of the normalized value $V_n = (V - V_{min})/(V_{max} - V_{min})$ where $V_{max}$ and $V_{min}$ are the maximum and minimum values of the $V$, as a function of the input intensity $I_0$ for the various frequencies of the input signal $\omega_s$ detuned from the resonant frequency $\omega_m$, so that $\Delta = \omega_s - \omega_m$. Here and after we use the bistability threshold intensity $I_{th}^{II}$ and the frequency shift threshold $\sigma_{th}$ for normalization.

The calculated diagrams may be explained as follows. For $I_0/I_{th}^{II} < 1$, the bistability is not observed and any cavity mode is stable regardless of the input signal detuning. In the case $I_0/I_{th}^{II} = 1$ the threshold appears for $\Delta/\sigma_{th} = -1$. It is shown in Fig. 4.a by the black open square and corresponds to the square mark in Fig. 3. For the higher values of the intensity $I_0/I_{th}^{II} > 1$, the two-valued range, where the signal detuning $\Delta$ leads to the three possible different values of the intracavity intensity, is observed. The stable values are given in Fig. 4.a by the black solid lines, while the black dashed lines represent the intermediate unstable states. The kinks of the nonlinear surface, shown in the same figure by the open orange and blue circles, define the instability band limits. These kinks correspond to the frequency positions of the maximum and minimum on the nonlinear dispersion characteristic of the mode with index $m$ (see the blue and orange open circles in Fig. 3), respectively. Blue and orange solid lines that connect the kink points on the nonlinear surface were obtained by substitution of the solution of Eq. (10) into Eq. (9). The projections of these lines onto a plane $(I_0/I_{th}^{II}, \Delta/\sigma_{th})$ in Fig. 4.a define the bounds to an area of the bistable behavior. These bounds represent the dynamic range of the bistability that are shown in the figure bottom plane by $I_m^+$ and $I_m^-$ (see the blue and orange dashed lines in Fig. 4.a). Further increase of the input intensity leads to extending of the multi-valued range. In case this range overlaps with the bistable zone of the neighboring mode with the index $m-1$ in a way that $I_m^+ = I_{m-1}^-$, the *tristable* region is formed (see Fig 4.b). The threshold intensity $I_{th}^{III}$, which is required to manifest the tristability can be found from Eq. (9) and Eq. (10) written for two neighboring modes. For the illustrative purposes, the bistable regions bounded by orange dashed lines ($I_m^+$ and $I_m^-$) and by green dashed lines ($I_{m-1}^+$ and $I_{m-1}^-$) are shown in the bottom plane of Fig. 4.b by the areas hatched by orange and green solid lines, respectively. The tristability appears as the intersection region of two bistable responses (see the cross hatching area by the both green and orange solid lines in the bottom plane in Fig. 4.b). Note that in the single RC, tristability phenomenon manifests itself due to the overlapping of two nonlinear responses of the neighboring resonant modes. Below the tristability threshold, the modes are in no way connected with each other, nor are the bistability effects manifested themselves at the neighboring resonant frequencies. However, the imposition of these isolated bistabilities leads to formation of the multistable behavior.

As was mentioned in the introduction, the multistability threshold can be significantly reduced in a system of the coupled RCs. This effect will be considered in the next paragraph.

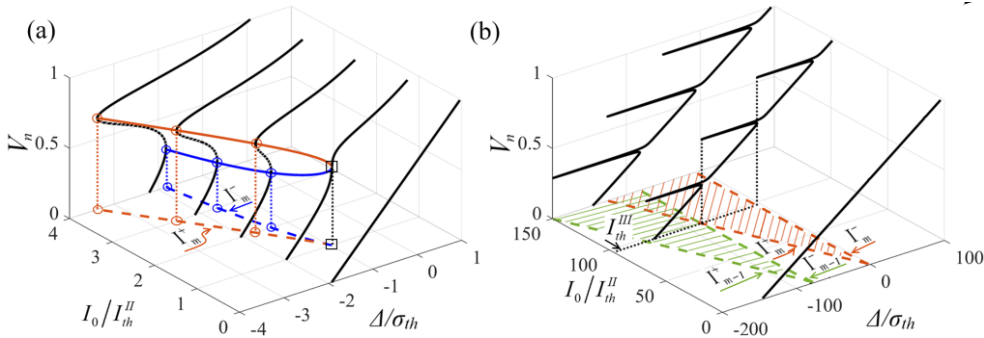

Fig. 4. Nonlinear surfaces of the normalized value $V_n$ as a function of the normalized intensity and frequency detuning showing bistability (a) and tristability regimes (b).

## 6. Multistability phenomenon in multi-ring resonant system

Below we investigate a multistability behavior of a multi-ring resonant circuit consisting of $N$ rings. For this purpose, the developed theoretical approach was extended for the case shown in Fig. 5, where $\kappa_{1i}$ and $\kappa_{2i}$ are the drop-in and drop-out power coupling coefficients for the RC with number $i$. Such design is used to demonstrate the capabilities of our theoretical approach and to describe the formation of multistable response, similar to presented in the previous Section for a single RC, but for a lower input power level. Moreover, the dynamics of the proposed circuit is simple and intuitive in comparison with more complex systems of coupled rings that are described in the works [10-12].

To simplify the analysis, we assume that the pumping signal $A_{in}$ is splitted between RCs in such a way that the phases and amplitudes of each signal at the RC inputs are equal. In other words, we assume that the input intensity of the signal into each RC is $I_{in} = I_0/N$. Such multi-ring circuit permits the signals to circulate in the rings independently, accumulate intracavity intensity $I_{c_i}$, and constructively interfere at the output port producing $I_{out} = |A_{out}|^2$. Therefore, the frequency response of the circuit is given as a sum of the frequency responses of each ring obtained with Eq. (3). According to the developed algorithm, one can calculate the nonlinear response surface for the resonant circuit with an arbitrary number of RCs.

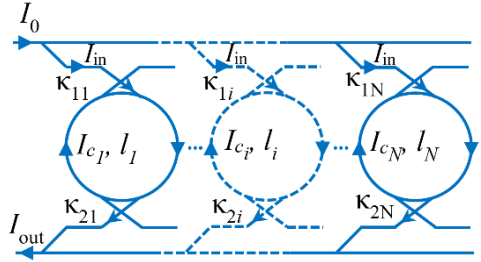

Fig. 5. Schematic of the multi-ring resonant circuit.

In order to demonstrate the multistability phenomenon we choose the two-RC configuration with slightly different ring lengths $l_1$ and $l_2$, which provides a short frequency distance between two neighboring frequency modes with numbers $m_1$ and $m_2$ equal to $\omega_{m_1} - \omega_{m_2} = 5.375\sigma_{th}$. Fig. 6 shows the color-mapped images of the nonlinear response surfaces for the different viewing angles. The construction of the figure was carried out with the same notations as in Fig. 2 for the single RC.

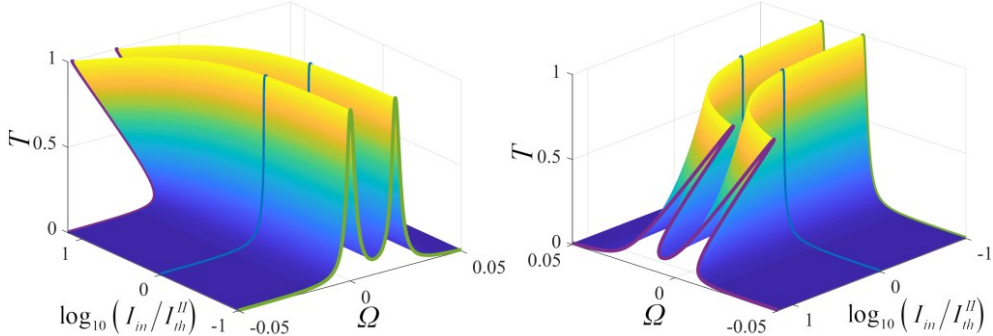

Fig. 6. Color-mapped nonlinear response surfaces for the two-RC resonant circuit calculated for the various intensities and shown for the different viewing angles.

In the beginning, so long as $I_{in} < I_{th}^{II}$, the frequency response of the circuit is almost linear (see green lines in Fig. 6). An increase of the input intensity provides an appearance of the bistability phenomena for the both RC (see blue lines in Fig. 6). The range of bistability extends with increasing of an input intensity. When $I_{in} \geq I_{th}^{III}$, the bistable responses overlap and the tristability phenomenon manifests itself as an appearance of the additional stable and unstable states (see magenta lines in Fig. 6). For identification of the bistability and tristability thresholds as well as for understanding the physical reasons for their formation, we carried out a theoretical analysis of the nonlinear diagram for the considered resonant circuit.

The obtained diagram is presented in Fig. 7.a. It shows a dynamic range of the multistable phenomena. The modeling results may be explained in the same manner as in previous section devoted to the bistability dynamic range. Here the black solid and dashed lines show the stable and unstable values, respectively. The kinks of the nonlinear surface constitute the blue and orange solid lines as well as the magenta and green solid lines showing the nonlinear behavior of mode with index $m_1$ for the first RC and index $m_2$ for the second one, respectively. The projections of these lines onto a plane $\left(I_0/I_{th}^{II},\Delta/\sigma_{th}\right)$ in Fig. 7.a show the bistable zones within the boundaries $I_{m_1}^+,I_{m_1}^-$ (see the orange and blue dashed lines) and $I_{m_2}^+,I_{m_2}^-$ (see the magenta and green dashed lines). The bistable regions are shown in the bottom plane of Fig. 7.a by the areas hatched by the orange and green solid lines, respectively. The tristability appears at $I_0 = I_{th}^{III}$ where the condition $I_{m_1}^- = I_{m_2}^+$ is satisfied. As is seen from Fig. 7.a, the tristability threshold decreases by a factor of 10 in comparison with Fig. 4.b. Further increasing of the input intensity provides broadening of the tristability region limited by the boundaries $I_{m_1}^-,I_{m_2}^+$ (see the cross hatching area by the both green and orange solid lines in the bottom plane of Fig. 7.a). The tristability condition and its bounds coincide numerically with the results described by Dumeige [11].

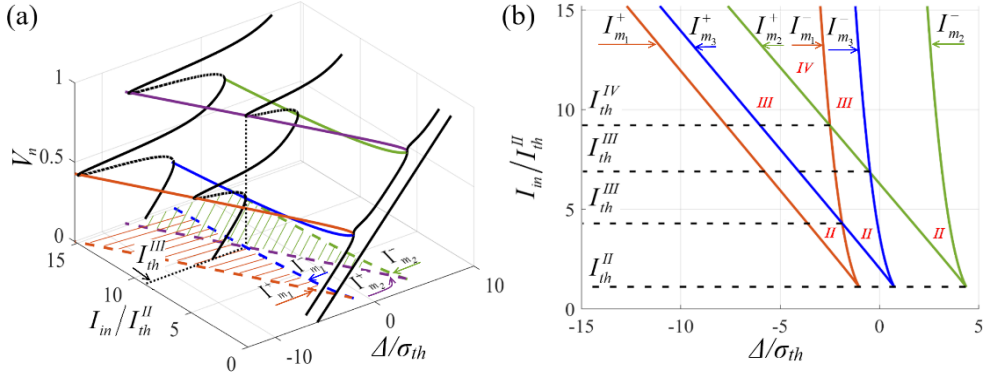

Fig. 7. Nonlinear surfaces of the normalized value $V_n$ as a function of the normalized intensity and frequency detuning showing the multistability in the two-RC circuit (a) and projection of this surface onto a plane $\left(I_0/I_{th}^{II},\Delta/\sigma_{th}\right)$ showing the multistability in the three-RC circuit (b).

A subsequent increase in the input intensity leads to overlapping of the tristability regions, which provides an appearance of one more stable and unstable states or *quadristability* [38]. As in the previous case for the tristability, the quadristability threshold $I_{th}^{IV}$ can be significantly reduced by appending one more RC with the slightly different $l_3$ so that the distance between resonant frequencies is $\omega_{m_1} - \omega_{m_3} = 1.75\sigma_{th}$. The boundaries of the bistability regions for each RC with $l_1, l_2$, and $l_3$ are shown in Fig. 7.b by the orange $\left(I_{m_1}^+,I_{m_1}^-\right)$, the green $\left(I_{m_2}^+,I_{m_2}^-\right)$, and the blue $\left(I_{m_3}^+,I_{m_3}^-\right)$ lines, respectively. This figure shows that crossings of the orange and blue lines $I_{m_1}^- = I_{m_3}^+$ as well as the green and blue ones $I_{m_3}^- = I_{m_2}^+$ provide the tristability phenomenon. Quadristability threshold manifests itself as a crossing of the orange and green lines $I_{m_1}^- = I_{m_2}^+$, after which these lines determine the quadristability region.

## 7. Conclusion

A convenient for practical purposes theoretical approach uniting the dispersion law and the frequency response of the multi-resonant nonlinear ring cavities is proposed for the first time. The approach provides an opportunity to study the nonlinear wave processes in a wide frequency range, significantly exceeding the distance between the adjacent resonant RC frequencies. The proposed approach is a visual tool that enables one to study the dynamics of the nonlinear response of the single-ring and multi-ring resonant circuits. To use the approach, it is enough to know the linear dispersion law and the dependence of material parameters of the ring on the wave amplitude. As an illustration for using, the main characteristics of the bistability and multistability phenomena in an optical ring with the Kerr nonlinearity are considered. The existence of many-valuedness of stable and unstable values of the intracavity intensity as well as the wave-numbers is shown. The physical mechanism of an appearance of the multistability phenomena and techniques to reduce their thresholds are demonstrated. It is shown that the new stable and unstable states can be progressively developed with increasing the number of RC in multi-ring resonant circuit. The developed approach can be extended to other more complex cases of the resonant ring circuits, which paves the way for the future research.

**Declaration of Competing Interest**

The authors declare that they have no known competing financial interests or personal relationships that could have appeared to influence the work reported in this paper.

**Acknowledgments**

This work was supported by the Ministry of Science and Higher Education of the Russian Federation (Project "Goszadanie" number 0788-2020-0005) and the Council on grants of the President of the Russian Federation (Grant #MK1938.2020.8).